\documentclass[3p,times,procedia]{elsarticle}
\usepackage{ecrc}
\usepackage{color}
\usepackage{bm}               

\usepackage{graphicx}
\usepackage{booktabs}
\usepackage{array}
\usepackage{hyperref}

\newcommand{\x}{{\bm x}}

\newcommand{\vpre}{{U_{\rm w, pre}}}
\newcommand{\vpost}{{U_{\rm w, post}}}
\usepackage{bm}
\usepackage{times}

\usepackage{balance}

\usepackage{epic,eepic}
\usepackage{graphics}
\usepackage{epsfig}
\usepackage{pifont}
\usepackage{subfigure}
\newcommand{\be}{\begin{equation}}
\newcommand{\ee}{\end{equation}}

\usepackage{color}
\usepackage{amssymb}
\usepackage{amsmath}
\usepackage{graphicx}
\usepackage{color}
\usepackage{hyperref}

\usepackage{amssymb}
\usepackage{enumerate}
\usepackage{amsthm,amsfonts,amsmath}
\usepackage{bm}
\usepackage{graphicx}
\usepackage{graphics,epsf,psfrag,pifont,dsfont,bbold}
\usepackage{siunitx}
\usepackage{color}
\usepackage{xspace}
\usepackage{accents}

\usepackage{amsmath}
\usepackage{bm}
\usepackage{hyperref}
\usepackage{amssymb}
\usepackage{natbib}
\usepackage{graphicx}

\volume{00}

\firstpage{1}

\journalname{Procedia IUTAM}

\runauth{Anupam Gupta}

\jid{piutam}

\usepackage{amssymb}

\usepackage[figuresright]{rotating}

\begin{document}

\begin{frontmatter}



\dochead{IUTAM Symposium on Multiphase flows with phase change: challenges and opportunities, Hyderabad, India (December 08 -- December 11, 2014)}

\title{Deformation and break-up of viscoelastic droplets Using Lattice Boltzmann Models}


\author[a]{Anupam Gupta} 
\author[a]{Mauro Sbragaglia}

\address[a]{Department of Physics and INFN, University of ``Tor Vergata'', Via della Ricerca Scientifica 1, 00133 Rome, Italy}

\begin{abstract}
We investigate the break-up of Newtonian/viscoelastic droplets in a viscoelastic/Newtonian matrix under the hydrodynamic conditions of a confined shear flow. Our numerical approach is based on a combination of Lattice-Boltzmann models (LBM) and Finite Difference (FD) schemes. LBM are used to model two immiscible fluids with variable viscosity ratio (i.e. the ratio of the droplet to matrix viscosity); FD schemes are used to model viscoelasticity, and the kinetics of the polymers is introduced using constitutive equations for viscoelastic fluids with finitely extensible non-linear elastic dumbbells with Peterlin's closure (FENE-P). We study both strongly and weakly confined cases to highlight the role of matrix and droplet viscoelasticity in changing the droplet dynamics after the startup of a shear flow. Simulations provide easy access to quantities such as droplet deformation and orientation and will be used to quantitatively predict the critical Capillary number at which the droplet breaks, the latter being strongly correlated to the formation of multiple neckings at break-up. This study complements our previous investigation on the role of droplet viscoelasticity (A. Gupta \& M. Sbragaglia, {\it Phys. Rev. E} {\bf 90}, 023305 (2014)), and is here further extended to the case of matrix viscoelasticity.
 \end{abstract}

\cortext[cor1]{Anupam Gupta. Tel.: +39-06-7259-4591 ; fax: +39-06-2023507.}
\end{frontmatter}

\email{agupta@roma2.infn.it University of ``Tor Vergata''}


\enlargethispage{-7mm}
\begin{nomenclature}
\begin{deflist}
\defitem{${\bm u}/ {\bm u}^{\prime}$}\defterm{Droplet/matrix velocity}
\defitem{$p/ p^{\prime}$}\defterm{Droplet/matrix pressure}
\defitem{$\eta_{M,f,d}$}\defterm{Dynamic shear viscosity (matrix (M), fluid solvent (f), droplet (d))}
\defitem{$\lambda$}\defterm{Viscosity ratio between the dispersed (droplet) and continuum (matrix) phase}
\defitem{$W, U_w, \dot{\gamma}=2U_w/W$}\defterm{Gap spacing, wall velocity, shear rate}
\defitem{$R$}\defterm{Droplet radius}
\defitem{${\cal C}$, $\tau_P$, $\eta_P$}\defterm{Polymer conformation tensor, polymer relaxation time, polymer viscosity}
\defitem{$f,L$}\defterm{FENE-P potential, maximum elongation of the polymers}
\defitem{$\mbox{Ca}$, $\mbox{De}$}\defterm{Capillary number, Deborah number}
\defitem{$\tau_{em}$}\defterm{Droplet emulsion time}
\end{deflist}
\end{nomenclature}

\section{Introduction}\label{main}

Emulsion properties are largely determined by their microstructure which can be tuned and designed for a huge variety of applications~\cite{Christophher07}. In particular, deformation and break-up  of dispersed droplets determine the emulsion rheology~\cite{Larson}. Droplet deformation and break-up in Newtonian fluids have been extensively studied in the literature~\cite{Taylor34,Grace,Stone}. The effect of an unconfined shear flow on droplets of one fluid suspended in another immiscible fluid was first considered long time ago by Taylor~\cite{Taylor34}: he estimated the largest stable droplet radius by balancing the surface stresses due to interfacial tension and viscous stresses due to shear. A dimensionless measure of this balance is provided by the Capillary number $\mbox{Ca} =\eta_M \dot{\gamma} R/\sigma$, where $\eta_M$ is the dynamic viscosity of the fluid matrix, $\dot{\gamma}$ the shear rate, $R$ the droplet radius at rest and $\sigma$ the surface tension. Break-up occurs at a critical Capillary number $\mbox{Ca}_{\mbox{\tiny{cr}}}$ when the viscous forces overcome the surface forces. The problem of droplet deformation and break-up under confined shear flow between two parallel plates has also been addressed in a series of theoretical and experimental papers (see~\cite{Shapira,Sibillo06,Vananroye07,Janssen10} and references therein). It was suggested that under confined conditions, a uniform shear flow can be exploited to generate quasi monodisperse emulsions by controlled break-up at near-critical conditions~\cite{Sibillo06,Renardy07}. The properties of confined droplets that contain viscoelastic components are less studied~\cite{Cardinaels09,Minale10,Cardinaels11} and the critical conditions for break-up have been rarely explored so far. Recent experiments suggest that viscoelasticity changes profoundly the critical Capillary numbers in confined conditions~\cite{Cardinaels11}. Complementing these kind of results with the help of numerical simulations would be of extreme interest. Simulations provide easier access to quantities such as droplet deformation and orientation as well as the velocity flow field and pressure field inside and outside the droplet. The goal of this paper is to use numerical simulations to characterize the idealized problem of a Newtonian/viscoelastic droplet subject to simple shear in a confined viscoelastic/Newtonian matrix. 

\section{Theoretical Model}

Our numerical approach is based on a hybrid combination of Lattice-Boltzmann models (LBM) and finite difference (FD) schemes, the former used to model two immiscible fluids with variable viscosity ratio, and the latter used to model viscoelasticity using the FENE-P constitutive equations. LBM have already been used to model droplet deformation problems~\cite{Xi99,VanDerSman08,Komrakovaa13,Liuetal12} and also viscoelastic flows~\cite{Onishi2,Malaspinas10}. The approach we use has already been studied and validated in a dedicated work~\cite{SbragagliaGuptaScagliarini}, where we have provided evidence that the model is able to capture quantitatively rheological properties of dilute suspensions as well as deformation and orientation of single droplets in confined shear flows. We just recall here the relevant continuum equations which are integrated in both the droplet (d) and the matrix (M) phases. In the droplet phase we integrate both the NS (Navier-Stokes) and FENE-P reference equations: 
\begin{eqnarray}\label{EQ}
\rho \left[ \partial_t \bm u + ({\bm u} \cdot {\bm \nabla}) \bm u \right] 
&=&  - {\bm \nabla}P + {\bm \nabla} \left(\eta_{A} ({\bm \nabla} {\bm u}+({\bm \nabla} {\bm u})^{T})\right)+
              \frac{\eta_P}{\tau_P}{\bm \nabla} \cdot [f(r_P){\bm {\bm {\mathcal C}}}];  
                                                 \label{NS}\\
\partial_t {\bm {\mathcal C}} + (\bm u \cdot {\bm \nabla}) {\bm {\mathcal C}}
&=& {\bm {\mathcal C}} \cdot ({\bm \nabla} {\bm u}) + 
                {({\bm \nabla} {\bm u})^T} \cdot {\bm {\mathcal C}} - 
                \frac{{f(r_P){\bm {\mathcal C}} }- {{\bm I}}}{\tau_P}.
                                                   \label{FENE}
\end{eqnarray}
Here, $\eta_A$ is the dynamic viscosity of the fluid, $\eta_P$ the viscosity parameter for the FENE-P solute, $\tau_P$ the polymer relaxation time, $\rho$ the solvent density, $P$ the solvent pressure, $({\bm \nabla} {\bm u})^T$ the transpose of $({\bm \nabla} {\bm u})$, ${\bm {\mathcal C}}$ the polymer-conformation tensor, ${\bm I}$ the identity tensor, $f(r_P)\equiv{(L^2 -3)/(L^2 - r_P^2)}$ the FENE-P potential that ensures finite extensibility, $ r_P \equiv \sqrt{Tr(\mathcal C)}$ and $L$ is the maximum possible extension of the polymers~\cite{bird}. In the outer matrix (M) phase (indicated with a prime), the corresponding equations are
\begin{eqnarray}\label{EQB}
\rho^{\prime} \left[ \partial_t \bm u^{\prime} + ({\bm u}^{\prime} \cdot {\bm \nabla}) \bm u^{\prime} \right] 
&=&  - {\bm \nabla}P^{\prime}+ {\bm \nabla} \left(\eta_{B} ({\bm \nabla} {\bm u}^{\prime}+({\bm \nabla} {\bm u}^{\prime})^{T})\right)+
              \frac{\eta^{\prime}_P}{\tau^{\prime}_P}{\bm \nabla} \cdot [f(r^{\prime}_P){\bm {\bm {\mathcal C}^{\prime}}}];  
                                                 \label{NSb}\\
\partial_t {\bm {\mathcal C}}^{\prime} + (\bm u^{\prime} \cdot {\bm \nabla}) {\bm {\mathcal C}^{\prime}}
&=& {\bm {\mathcal C}}^{\prime} \cdot ({\bm \nabla} {\bm u}^{\prime}) + 
                {({\bm \nabla} {\bm u}^{\prime})^{T}} \cdot {\bm {\mathcal C}^{\prime}} - 
                \frac{{f(r^{\prime}_P){\bm {\mathcal C}^{\prime}} }- {{\bm I}}}{\tau^{\prime}_P}.
                                                   \label{FENEb}
\end{eqnarray} 
with $\eta_{B}$ the solvent matrix shear viscosity. In all the cases, immiscibility between the droplet phase and the matrix phase is introduced using the so-called ``Shan-Chen'' model~\cite{SbragagliaGuptaScagliarini}. In all the numerical simulations presented in this paper, we work with unitary viscosity ratio, defined in terms of the total (solvent+polymer) shear viscosity. In particular, when studying matrix viscoelasticity (MV), we will choose a case with $\eta_P=0$ in equation (\ref{NS}) and $\lambda=\eta_d/\eta_M=\eta_A/(\eta_B+\eta^{\prime}_P)=1$ and polymer concentration $\eta^{\prime}_P/\eta_M \approx 0.4$; for the simulations with droplet viscoelasticity (DV), we will choose a case with $\eta^{\prime}_P=0$ in equation (\ref{NSb})  with $\lambda=\eta_d/\eta_M=(\eta_A+\eta_P)/\eta_B=1$ and polymer concentration $\eta_P/\eta_d \approx 0.4$. The degree of viscoelasticity is computed from the Deborah number 
\be\label{De}
\mbox{De}=\frac{N_1 R}{2 \sigma}\frac{1}{\mbox{\mbox{Ca}}^2}
\ee 
where $\mbox{Ca}$ is always computed in the matrix phase while the Deborah number is computed in either the matrix or the droplet phase, dependently on the case studied. In Eq. (\ref{De}), $N_1$ is the first normal stress difference which develops in homogeneous steady shear. Solving the constitutive equations for such a hydrodynamic problem, $u_x=\dot{\gamma} z$, $u_y=0$, $u_z=0$, both the polymer feedback stress and the first normal stress difference $N_1$ for the FENE-P model~\cite{bird,Lindner03} follow (primed variables replace non-primed variables for matrix phases)
\be\label{S}
\frac{\eta_P}{\tau_P} f(r_P) {\mathcal C}_{xz} =\frac{2 \eta_P}{\tau_P} \left(\frac{L^2}{6} \right)^{1/2} \sinh \left(\frac{1}{3} \mbox{arcsinh} \left(\frac{\dot{\gamma}\tau_P L^2}{4} \left(\frac{L^2}{6}\right)^{-3/2}\right) \right)
\ee
\be\label{N1}
N_1 = \frac{\eta_P}{\tau_P} f(r_P) ({\mathcal C}_{xx}-{\mathcal C}_{yy})=8 \frac{\eta_P}{\tau_P} \left(\frac{L^2}{6} \right) \sinh^2 \left(\frac{1}{3} \mbox{arcsinh} \left(\frac{\dot{\gamma} \tau_P L^2}{4} \left(\frac{L^2}{6}\right)^{-3/2}\right) \right).
\ee
In the Oldroyd-B limit ($L^2 \gg 1$) we can use the asymptotic expansion of the hyperbolic functions and we get  
\be\label{Desimple}
\mbox{De}=\frac{\tau_P}{\tau_{\mbox{\tiny{em}}}} \frac{\eta_P}{\eta_{M}}.
\ee
Equation (\ref{Desimple}) shows that $\mbox{De}$ is clearly dependent on the ratio between the polymer relaxation time $\tau_P$ and the droplet emulsion time 
\be\label{emulsiontime}
\tau_{\mbox{\tiny{em}}}=\frac{R \eta_{M}}{\sigma}. 
\ee
As evident from Eq. (\ref{S}), the model supports a thinning effect at large shear, although such effect will not be important in our calculations, all the numerical simulations being performed with fluid pairs with nearly constant shear viscosities. In the following sections, we report the Deborah number based on the definition \eqref{Desimple}, as we estimated the difference between~\eqref{Desimple} and~\eqref{De} to be at maximum of a few percent for the values of $L^2$ considered. Also, we focus mainly on the droplet deformation and break-up problems, being the quantitative benchmarks against known analytical results for the rheology of dilute suspensions~\cite{bird,Herrchen} present in another dedicated methodological publication~\cite{SbragagliaGuptaScagliarini}. In a previous study~\cite{SbragagliaGuptaPRE} we investigated the role of droplet viscoelasticity: a non trivial interplay between confinement and viscoelasticity has emerged. With the use of numerical simulations we had the opportunity to change separately the viscosity ratio of the Newtonian phases, the maximum extension of the polymers, and the degree of viscoelasticity, thus allowing for a systematic analysis of the viscoelastic effects while keeping the shear viscosity of the droplet fixed to the reference Newtonian case. In particular, by increasing the finite extensibility of the polymers, it was observed that the resistance against elongation may be enough to prevent both droplet elongation and subsequent triple break-up, thus altering significantly the critical Capillary number for viscoelastic droplets under confinement. In this paper, we push the analysis a bit further and we propose a comparative study between matrix and droplet viscoelasticity.  To simplify matters, we will also keep the maximum elongation of the polymers fixed to $L^2=100$, since we have exhaustively treated the importance of $L^2$ in our previous study~\cite{SbragagliaGuptaPRE}. 

\section{Results}


\begin{figure}[t!]
\begin{center}
\hspace{-1.0cm}
\includegraphics[scale=0.35]{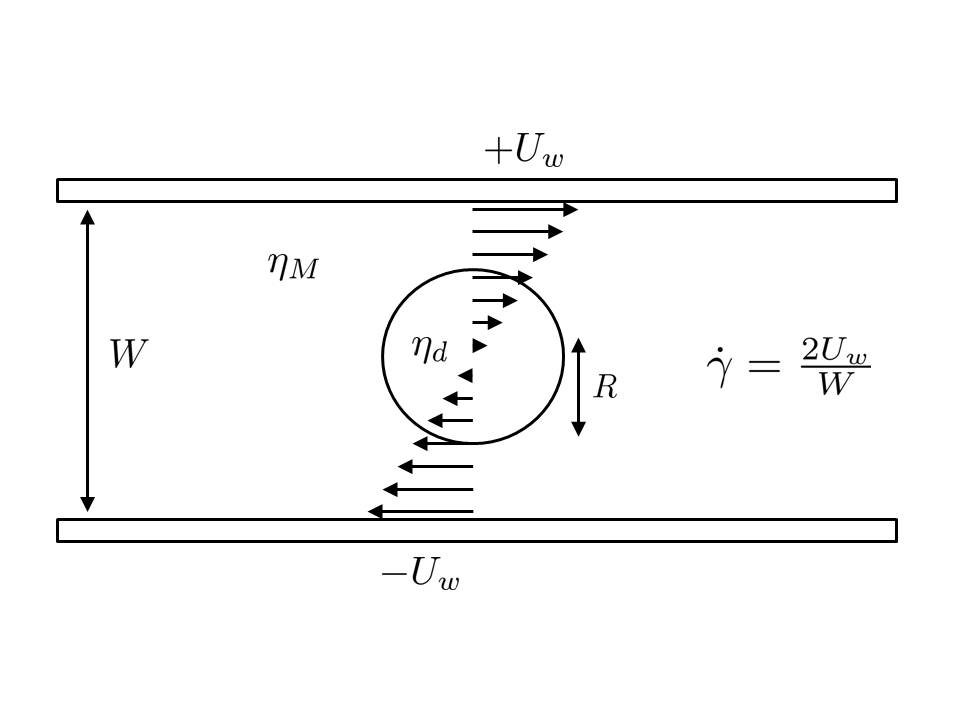}
\end{center}
\caption{Shear plane view of the numerical set-up for the study of break-up of confined droplets. A Newtonian droplet (Phase $A$) with radius $R$ and shear viscosity $\eta_A$ is placed in between two parallel plates at distance $W$ in a Newtonian matrix (Phase $B$) with shear viscosity $\eta_B$. A polymer phase with shear viscosity $\eta_P$ ($\eta^{\prime}_P$) is then added in the droplet (matrix) phase. The kinetics of the polymers is introduced using constitutive equations for viscoelastic fluids with finitely extensible non-linear elastic dumbbells with Peterlin's closure (FENE-P). The equation set is summarized in Eqs. \eqref{EQ}-\eqref{FENEb}. We work with unitary viscosity ratio, defined in terms of the total (fluid+polymer) shear viscosity: $\lambda=(\eta_A+\eta_P)/\eta_B=1$, in case of droplet viscoelasticity; $\lambda=\eta_A/(\eta_B+\eta^{\prime}_P)=1$, in case of matrix viscoelasticity. A shear is applied by moving the two plates in opposite directions with velocities $\pm U_w$.}
\label{fig:sketch}
\end{figure}

\begin{figure}[t!]
\subfigure[t/$\tau_{\mbox{\tiny{em}}}$=25, 2R/W=0.52, De=0, Ca=0.34]
    {
        \includegraphics[scale=0.05]{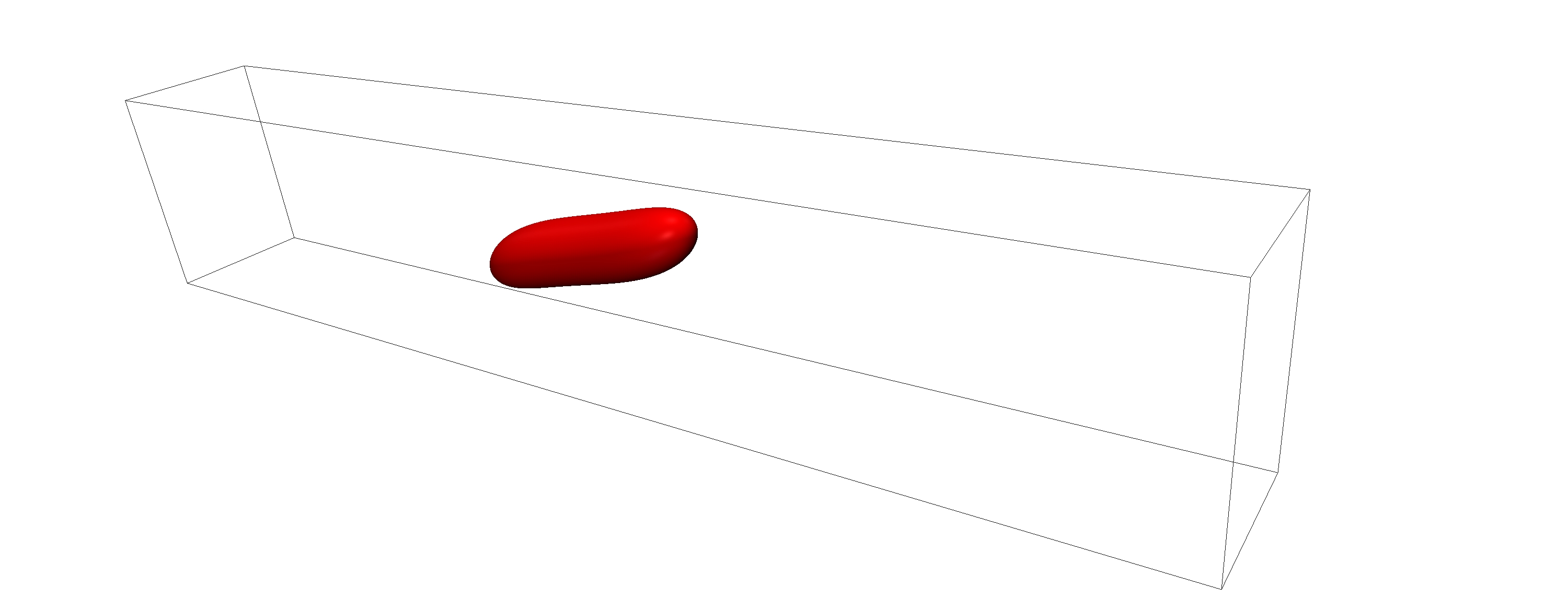}
    }    
\subfigure[t/$\tau_{\mbox{\tiny{em}}}$=75, 2R/W=0.52, De=0, Ca=0.34]
    {
        \includegraphics[scale=0.05]{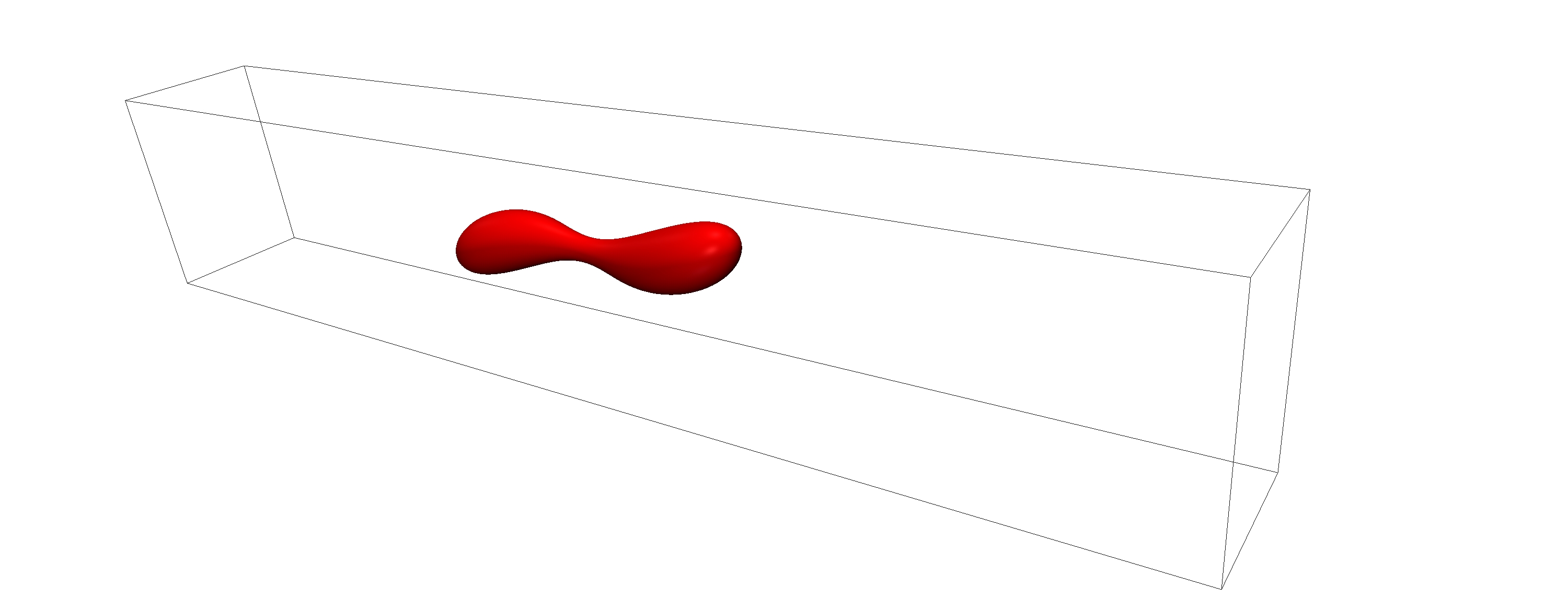}
    }
\subfigure[t/$\tau_{\mbox{\tiny{em}}}$=100, 2R/W=0.52, De=0, Ca=0.34]
    {
        \includegraphics[scale=0.05]{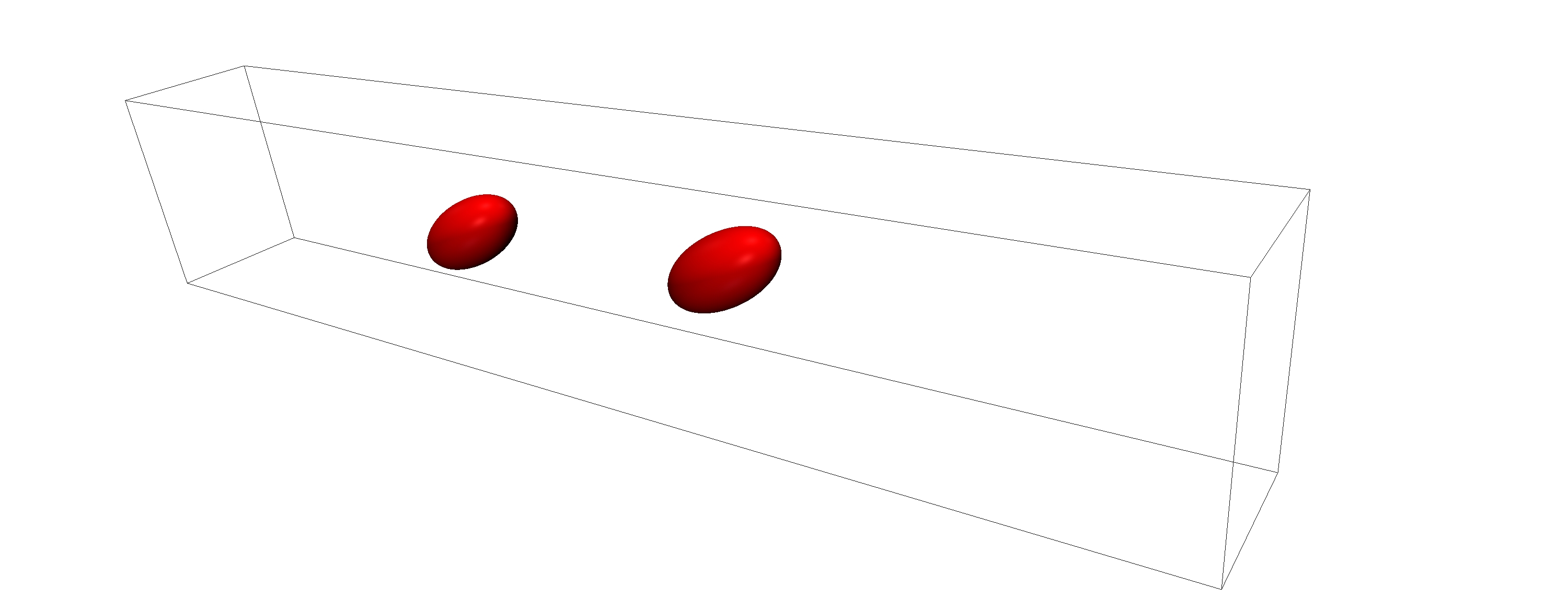}
    }    
\\
\subfigure[t/$\tau_{\mbox{\tiny{em}}}$=25, 2R/W=0.52, Ca=0.397 (DV)]
    {
        \includegraphics[scale=0.05]{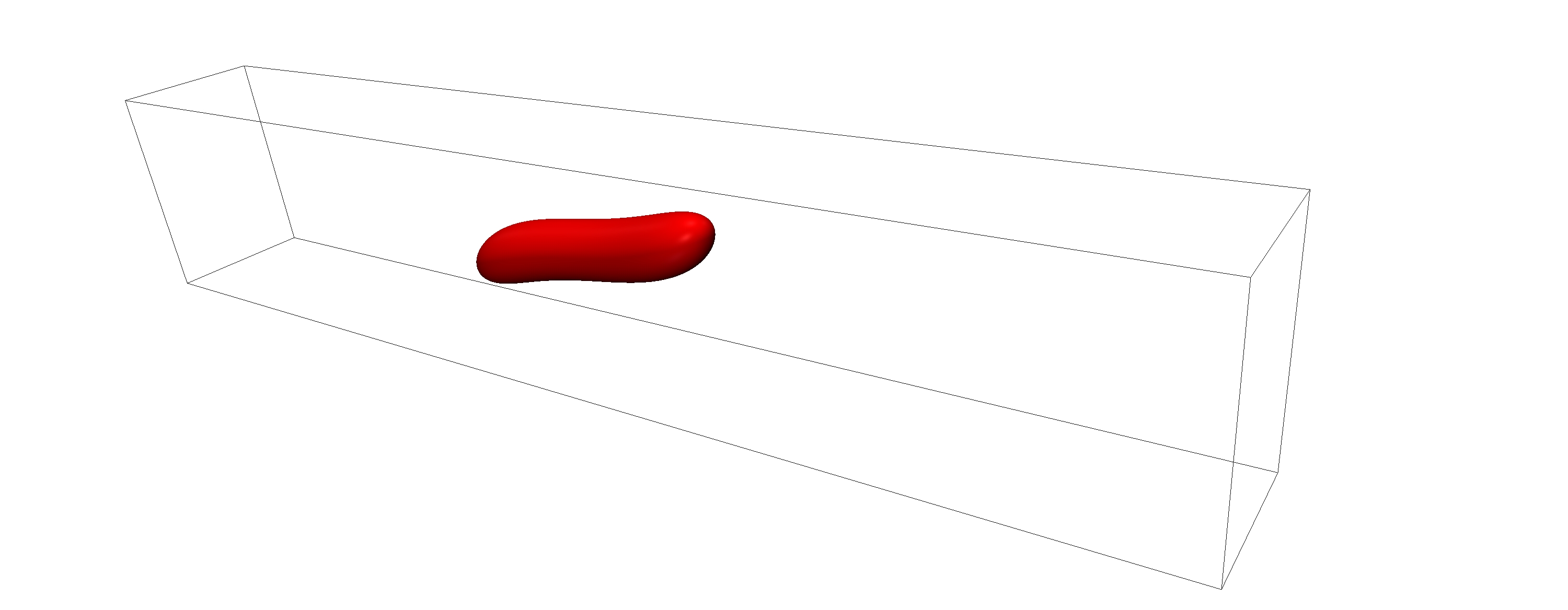}
    }    
\subfigure[\,\,t/$\tau_{\mbox{\tiny{em}}}$=75, 2R/W=0.52, Ca=0.397 (DV)]
    {
        \includegraphics[scale=0.05]{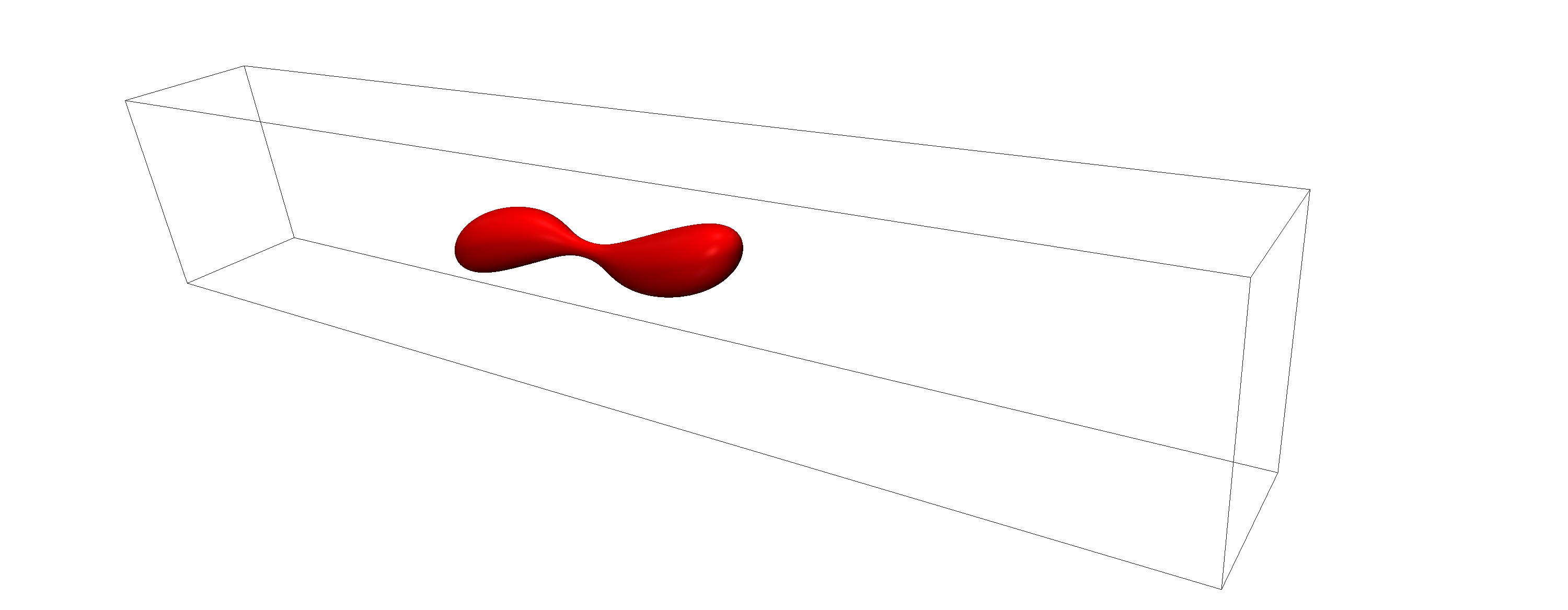}
    }    
\subfigure[\,\,t/$\tau_{\mbox{\tiny{em}}}$=100, 2R/W=0.52, Ca=0.397 (DV)]
    {
        \includegraphics[scale=0.05]{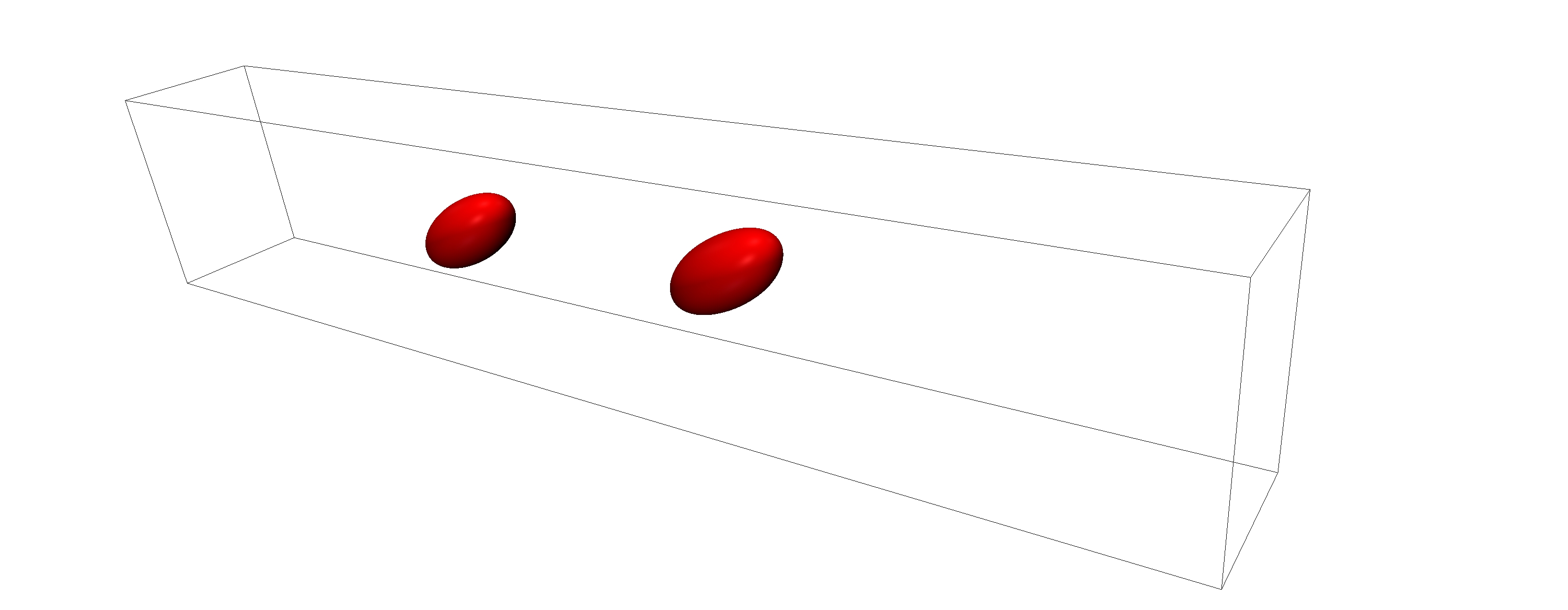}
    }
\\
\subfigure[\,\,t/$\tau_{\mbox{\tiny{em}}}$=25, 2R/W=0.52, Ca=0.34 (MV)]
    {
        \includegraphics[scale=0.05]{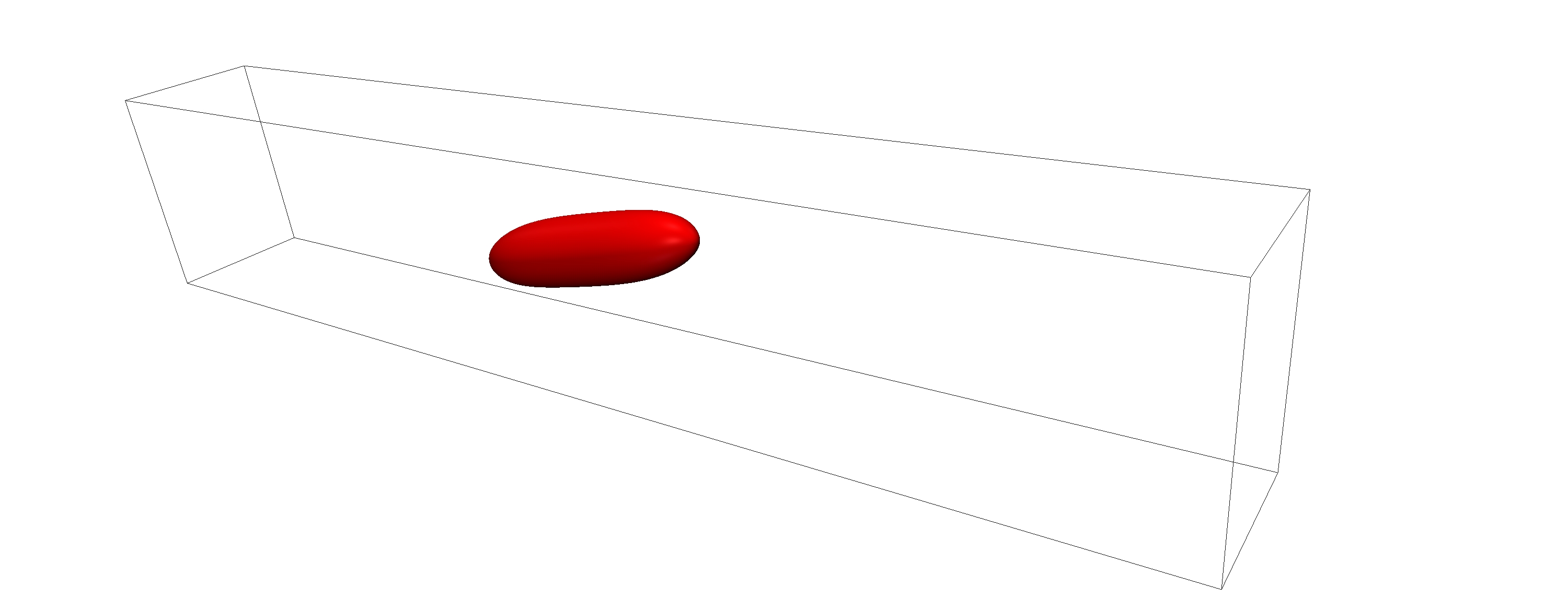}
    }    
\subfigure[\,\,t/$\tau_{\mbox{\tiny{em}}}$=75, 2R/W=0.52, Ca=0.34 (MV)]
    {
        \includegraphics[scale=0.05]{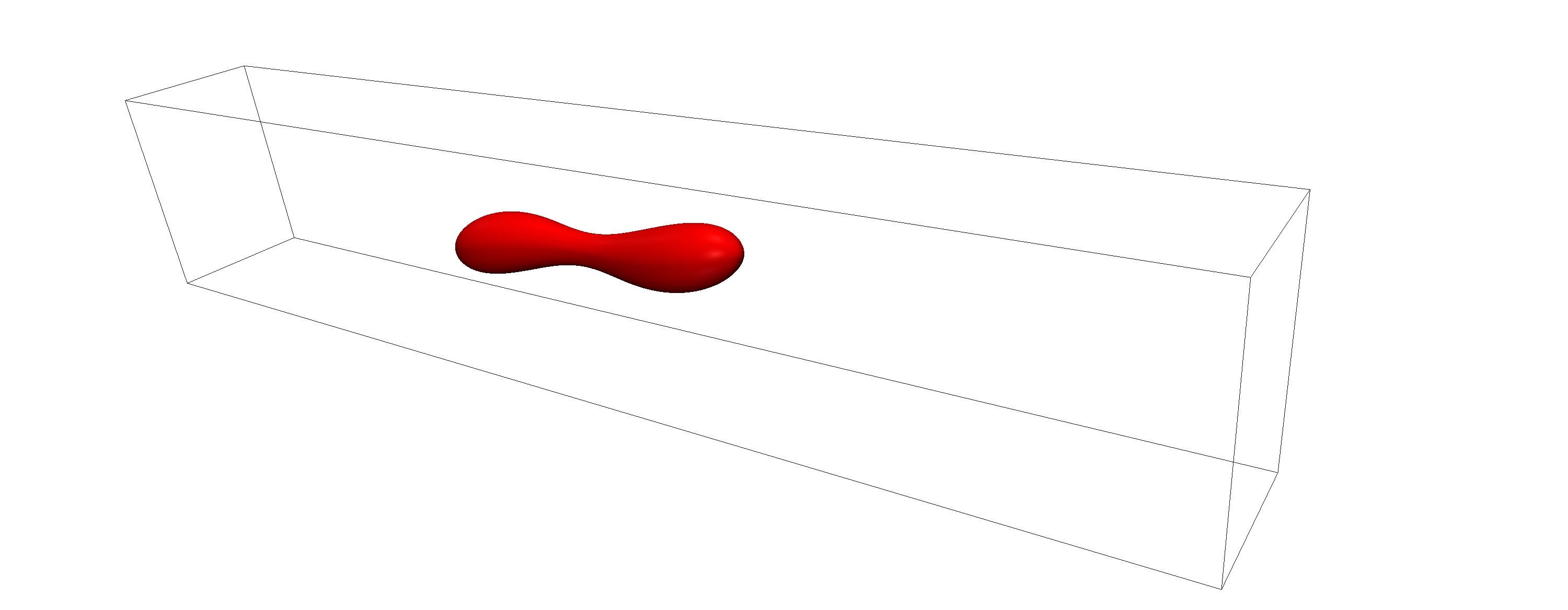}
    }    
\subfigure[\,\,t/$\tau_{\mbox{\tiny{em}}}$=100, 2R/W=0.52, Ca=0.34 (MV)]
    {
        \includegraphics[scale=0.05]{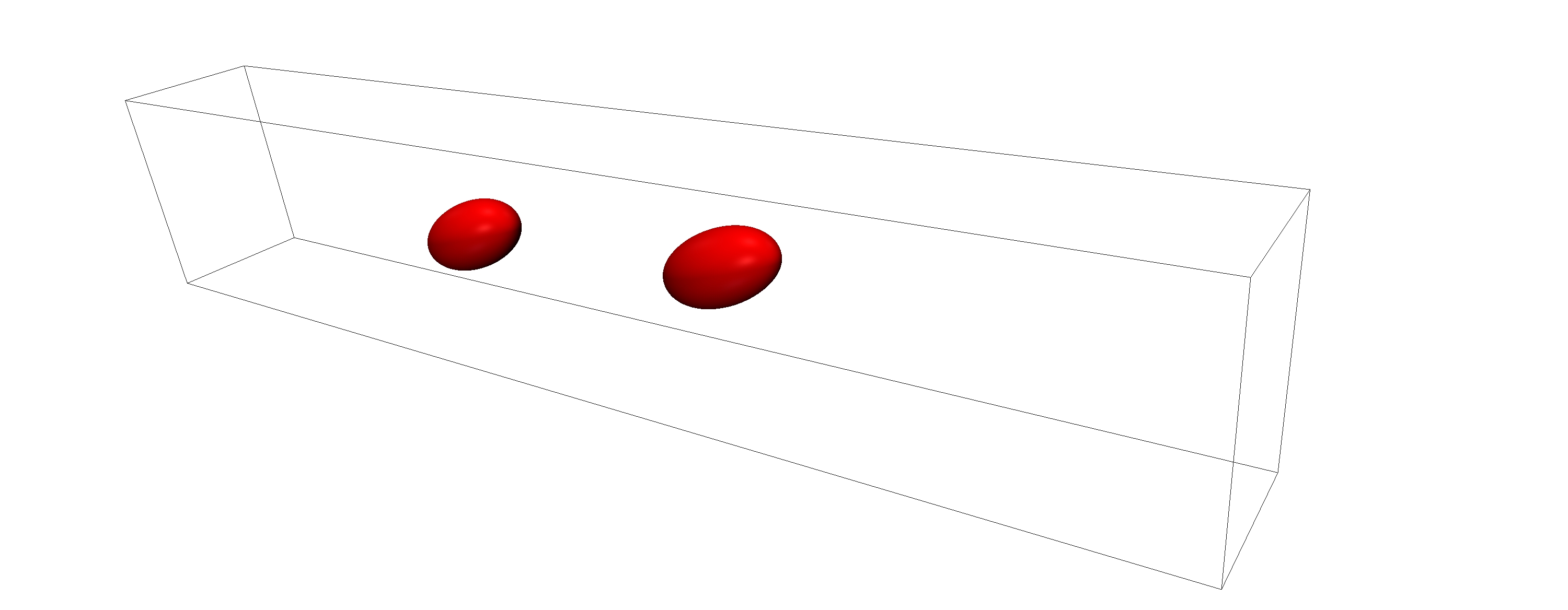}
    }
\\
\caption{Break-up after the startup of a shear flow with confinement ratio $2R/W=0.52$. We report the time history of droplet deformation and break-up including 3 representative time frames: initial deformation (left column); deformation prior to break-up (middle column); post break-up frame (right column). We use the droplet emulsion time $\tau_{em}$ \eqref{emulsiontime} as a unit of time. Panels (a)-(c): Newtonian droplet in a Newtonian matrix. Panels (d)-(f): viscoelastic droplet (DV) with Deborah number $\mbox{De}=2.0$ in a Newtonian matrix. Non-Newtonian properties stabilize (very little) the droplet deformation with a slightly larger critical Capillary number, $\mbox{Ca}_{\mbox{\tiny{cr}}}=0.397$. Panels (g)-(i): Newtonian Droplet in a viscoelastic matrix (MV) with Deborah number $\mbox{De}=2.0$. Matrix viscoelasticity has an almost insignificant effect. In all cases, the viscosity ratio between the droplet phase and the matrix phase is kept fixed to $\lambda=\eta_d/\eta_M=1$, independently of the degree of viscoelasticity. \label{fig:1}}
\end{figure}
\begin{figure}[t!]
\subfigure[\,\,t/$\tau_{em}$=25, 2R/W=0.7, De=0, Ca=0.426]
    {
        \includegraphics[scale=0.05]{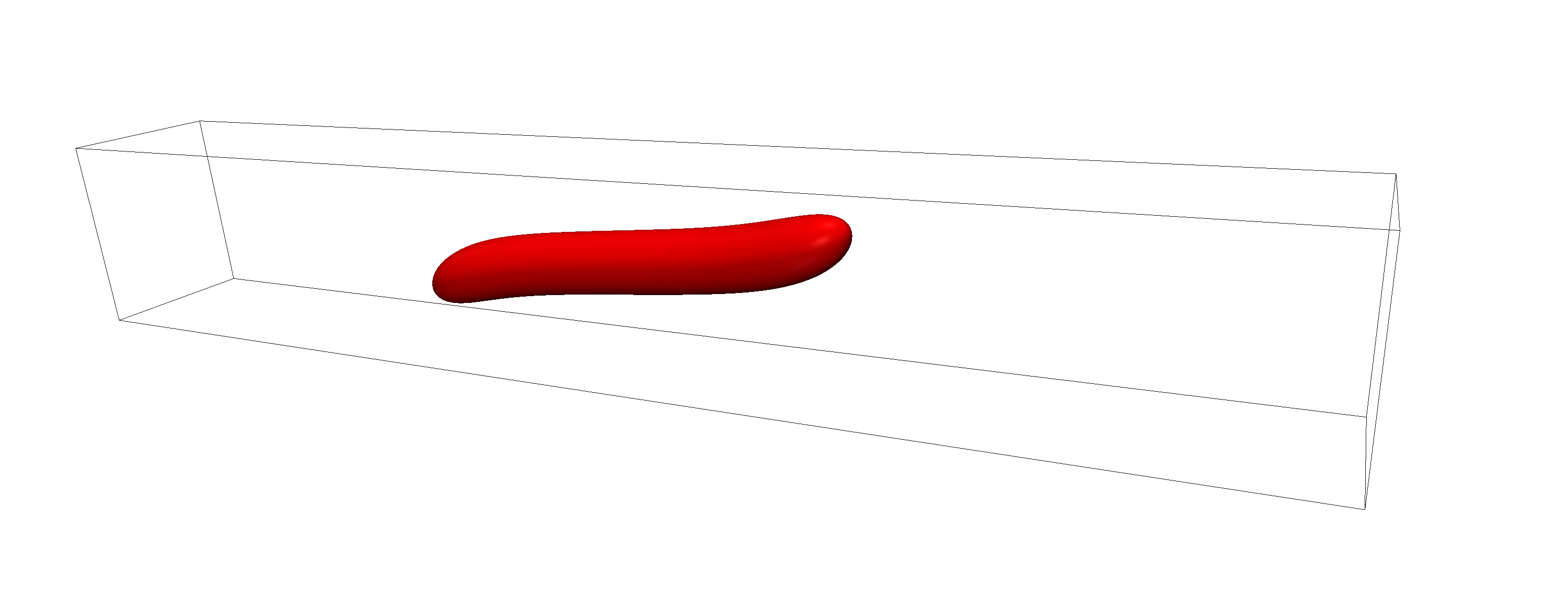}
    }    
\subfigure[\,\,t/$\tau_{em}$=75, 2R/W=0.7, De=0, Ca=0.426]
    {
        \includegraphics[scale=0.05]{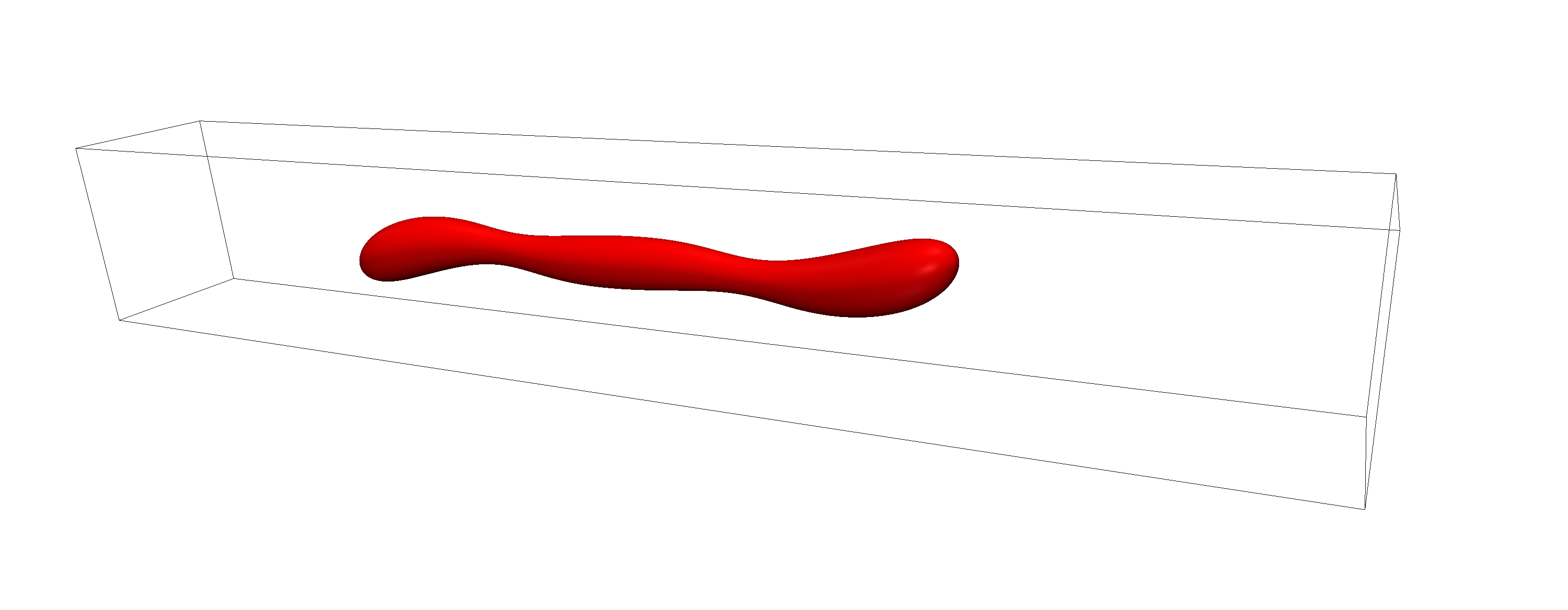}
    }
\subfigure[\,\,t/$\tau_{em}$=100, 2R/W=0.7, De=0, Ca=0.426]
    {
        \includegraphics[scale=0.05]{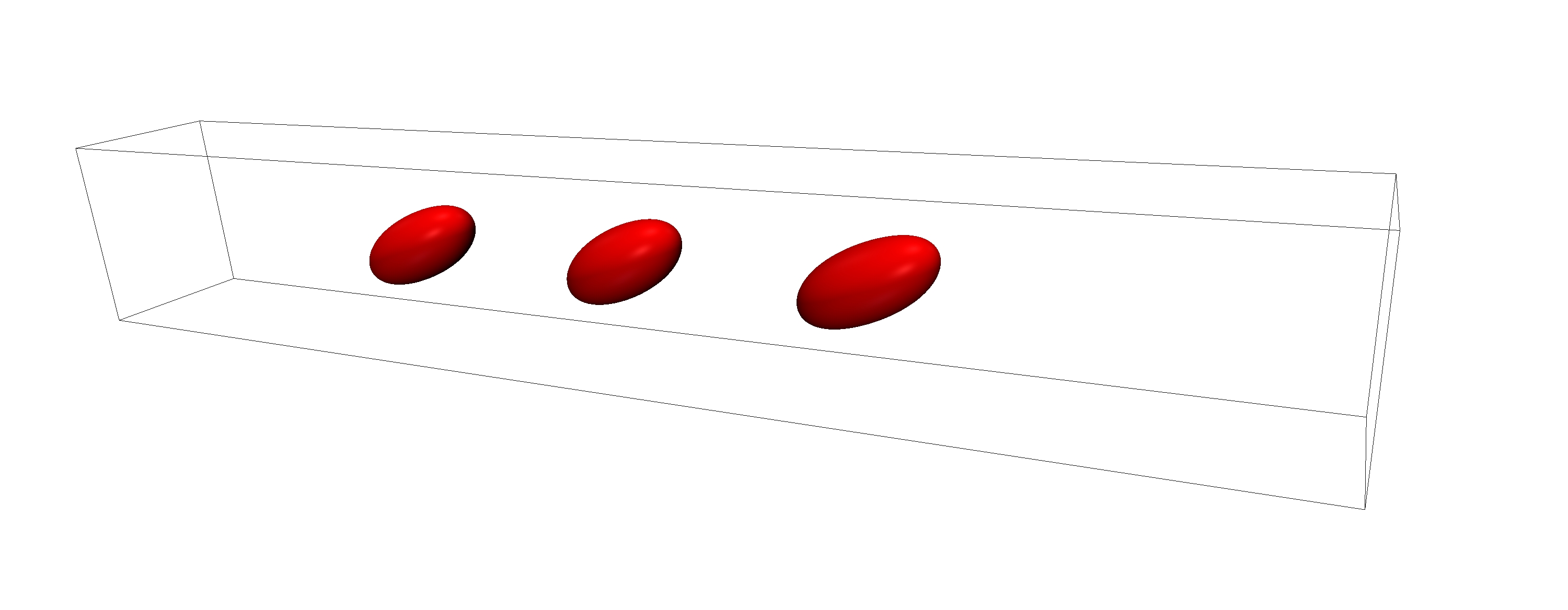}

    }    
\\
\subfigure[\,\,t/$\tau_{em}$=25, 2R/W=0.7, Ca=0.64 (DV)]
    {
        \includegraphics[scale=0.05]{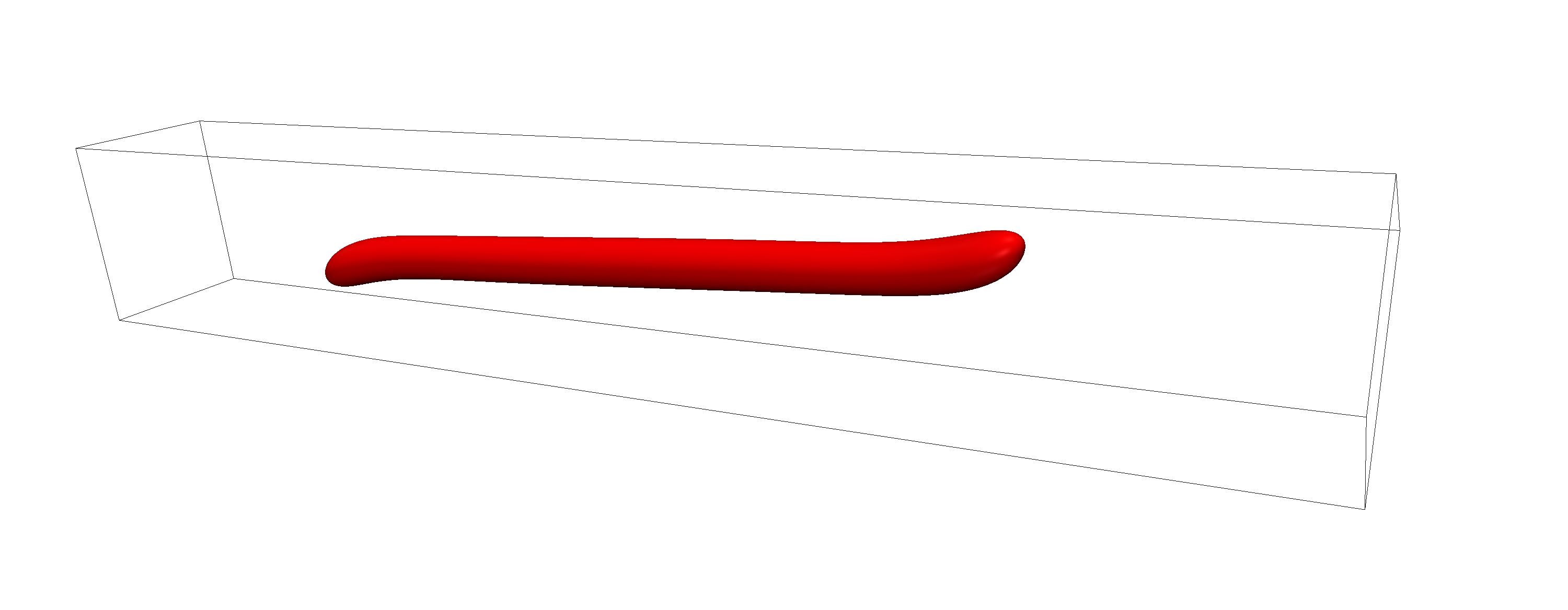}
    }    
\subfigure[\,\,t/$\tau_{em}$=75, 2R/W=0.7, Ca=0.64 (DV)]
    {
        \includegraphics[scale=0.05]{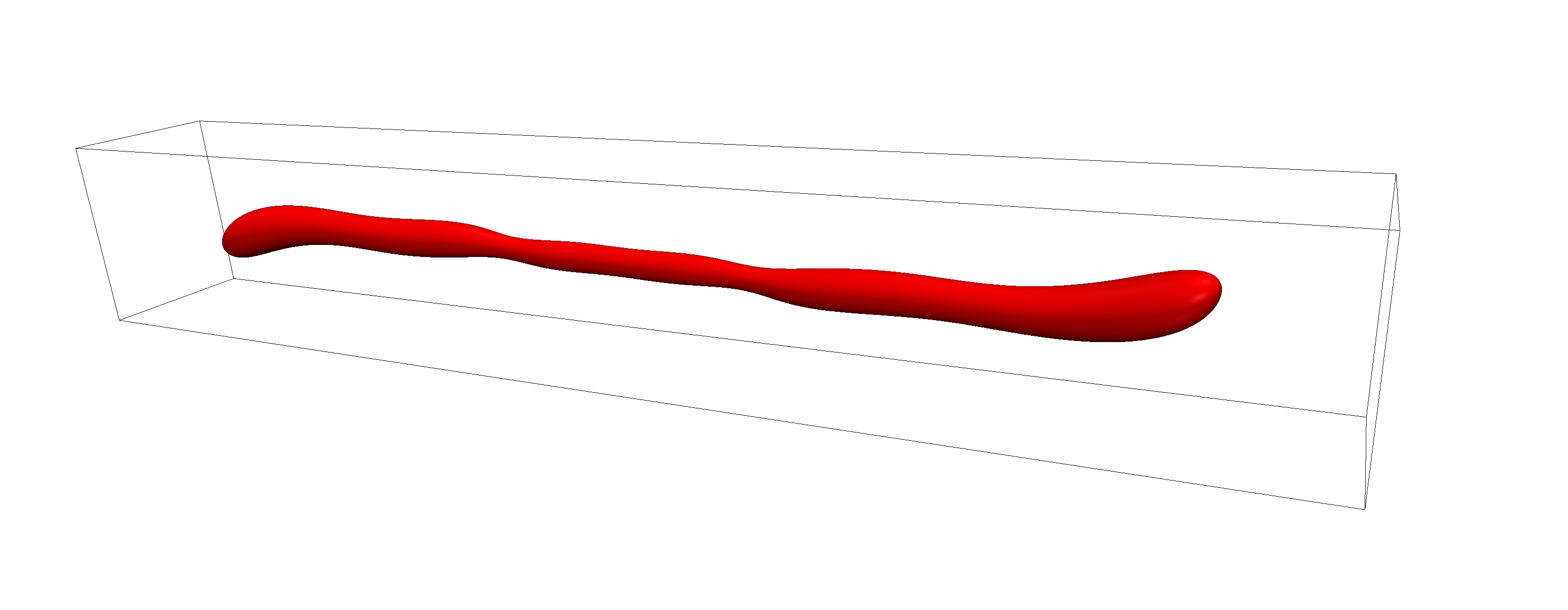}
    }
\subfigure[\,\,t/$\tau_{em}$=100, 2R/W=0.7, Ca=0.64 (DV)]
    {
        \includegraphics[scale=0.05]{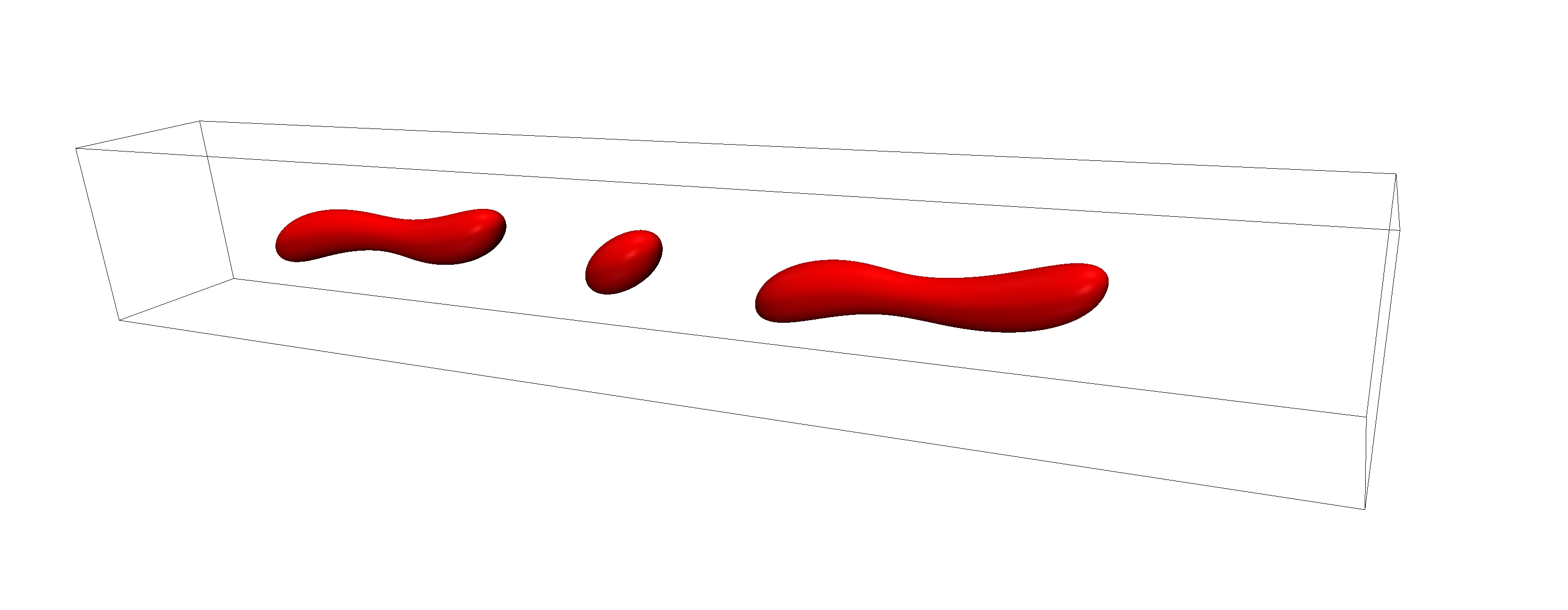}
    }
\\
\subfigure[\,\,t/$\tau_{em}$=25, 2R/W=0.7, Ca=0.32 (MV)]
    {
        \includegraphics[scale=0.05]{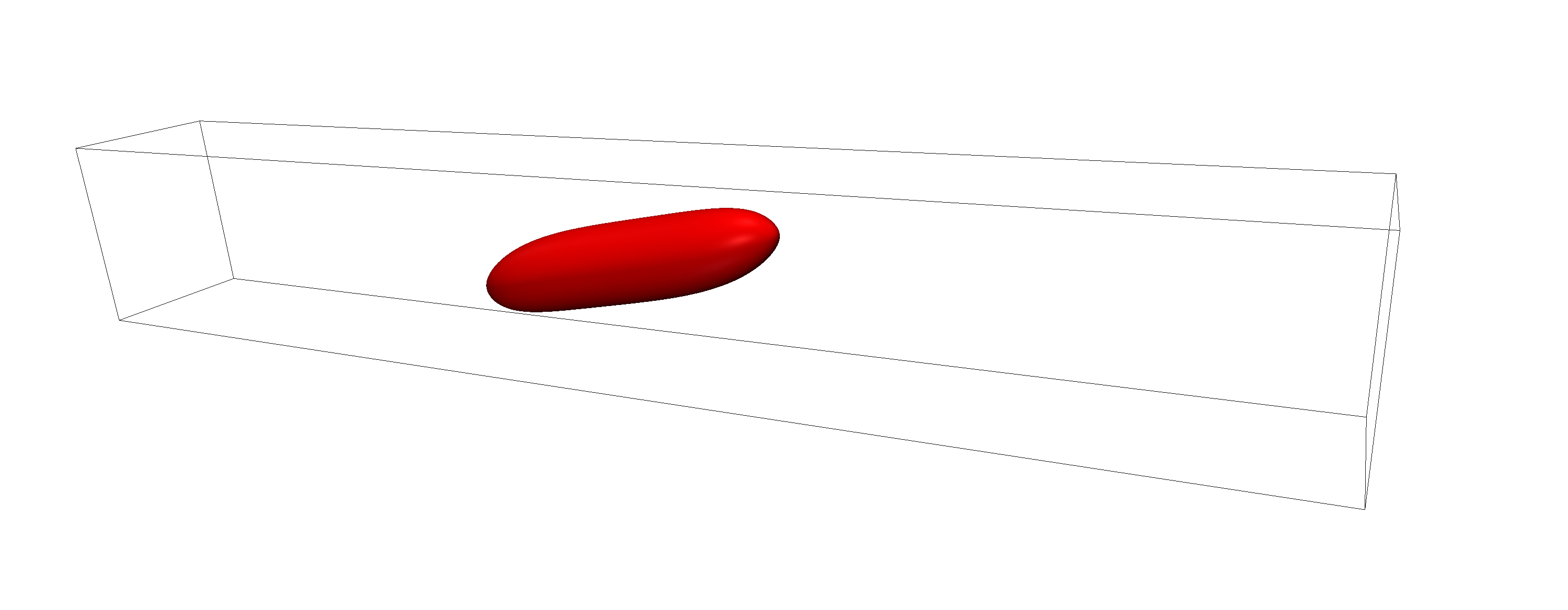}
    }    
\subfigure[\,\,t/$\tau_{em}$=75, 2R/W=0.7, Ca=0.32 (MV)]
    {
        \includegraphics[scale=0.05]{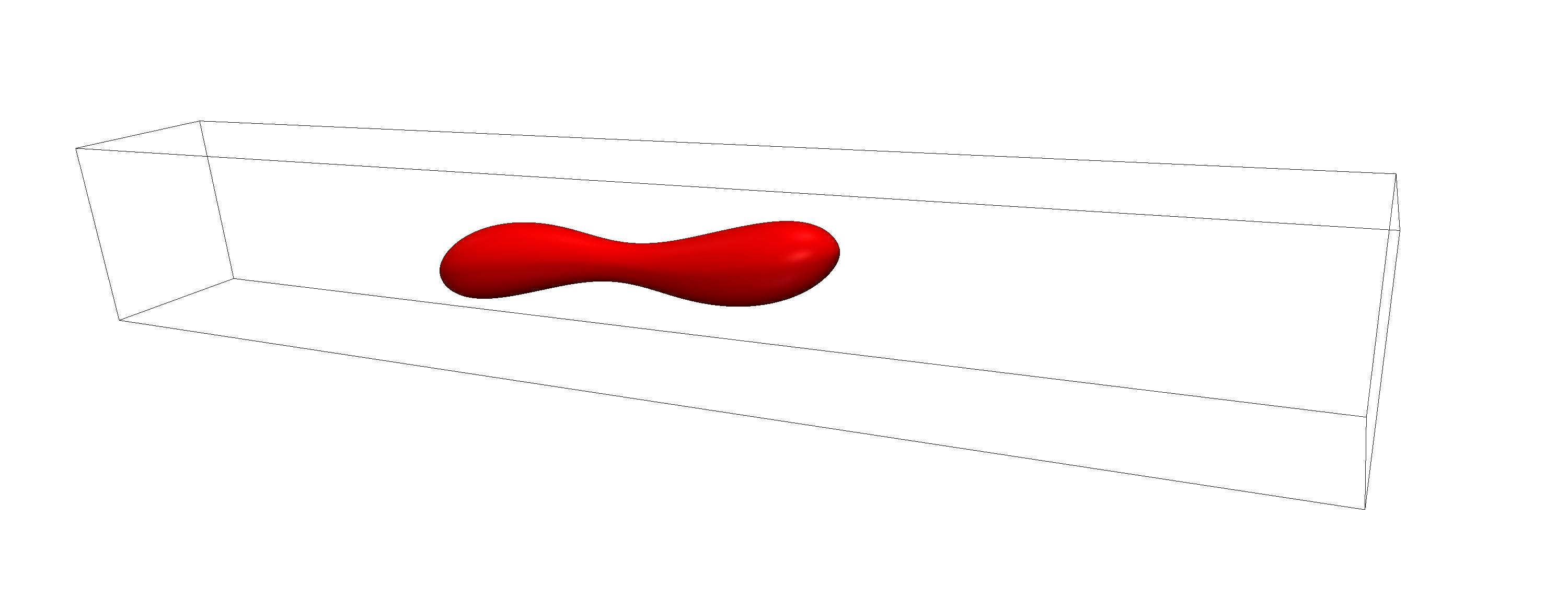}
    }
\subfigure[\,\,t/$\tau_{em}$=100, 2R/W=0.7, Ca=0.32 (MV)]
    {
        \includegraphics[scale=0.05]{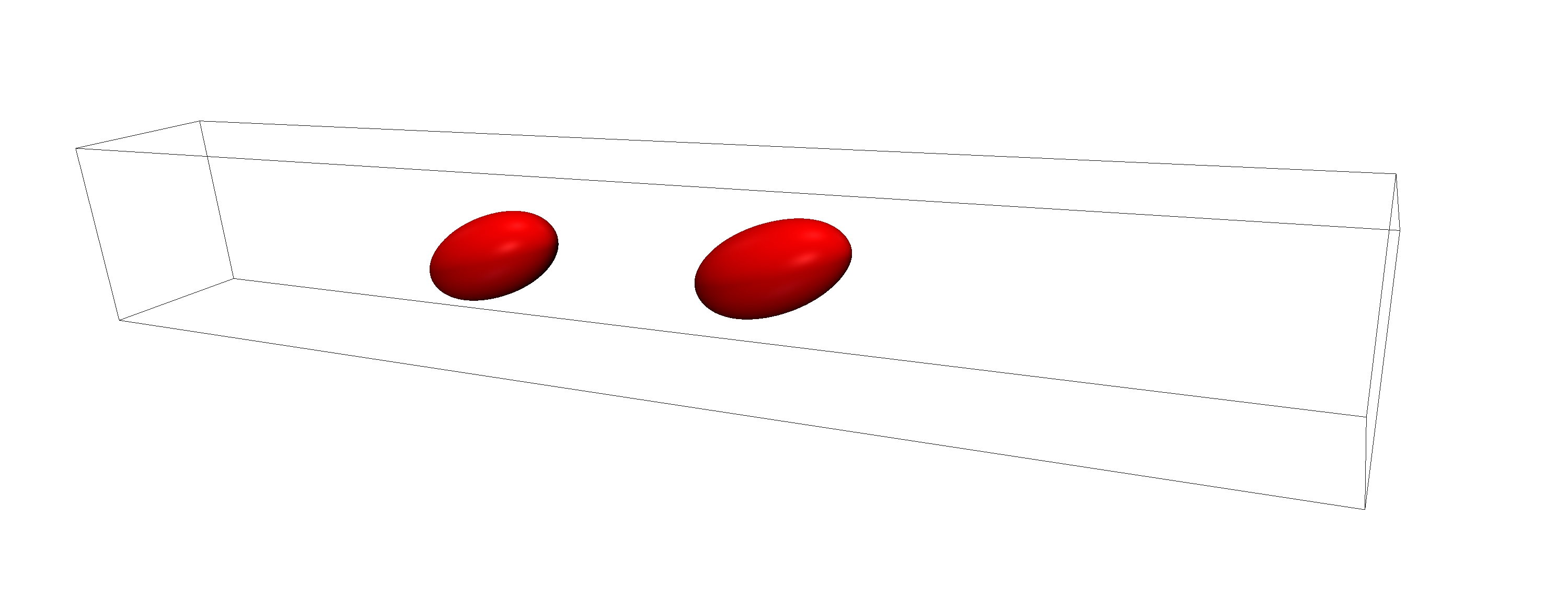}
    }
\\
\caption{Break-up after the startup of a shear flow with confinement ratio $2R/W=0.7$. We report the time history of droplet deformation and break-up including 3 representative time frames: initial deformation (left column); deformation prior to break-up (middle column); post break-up frame (right column). We use the droplet emulsion time $\tau_{em}$ \eqref{emulsiontime} as a unit of time.  Panels (a)-(c): Newtonian droplet in a Newtonian matrix.  A distinctive feature of this confined case is the emergence of {\it triple break-up}~\cite{Janssen10}. The critical Capillary number is estimated to be $\mbox{Ca}_{\mbox{\tiny{cr}}}=0.426$. Panels (d)-(f): viscoelastic droplet (DV) with Deborah number $\mbox{De}=2.0$ in a Newtonian matrix. Droplet viscoelasticity stabilizes the droplet deformation and inhibits droplet break-up. The critical Capillary number is estimated to be $\mbox{Ca}_{\mbox{\tiny{cr}}}=0.64$. Panels (g)-(i): Newtonian droplet in a viscoelastic matrix (MV) with Deborah number $\mbox{De}=2.0$. Matrix viscoelasticity destabilizes the formation of long droplet shapes, and the critical Capillary number, $\mbox{Ca}_{\mbox{\tiny{cr}}}=0.32$, is very similar to the unbounded case (see Fig.~\ref{fig:1}). In all cases, the viscosity ratio between the droplet phase and the matrix phase is kept fixed to $\lambda=\eta_d/\eta_M=1$, independently of the degree of viscoelasticity. \label{fig:2}}
\end{figure}

In all the cases discussed in this section, a spherical droplet with radius $R$ is initially placed halfway  between the walls. The critical Capillary number is computed by identifying the pre-critical  ($\vpre$) and the post-critical wall velocity ($\vpost$), i.e. the largest (smallest) wall velocity for which the droplet is stable (breaks). All the simulations described refer  to  cases with polymer relaxation times ranging in the interval $0 \le \tau_{P} \le 5000$ lbu and finite extensibility parameter $L^2= 10^2$, corresponding to Deborah numbers ranging in the interval $0 \le \mbox{De} \le 2$. The numerical simulations have been carried out in three dimensional domains  $L_{x} \times L_y \times W$. The droplet radius $R$ and the vertical gap $W$ have been changed in the ranges $50 \le R \le 60 $ lattice cells and $128 \le W \le 256 $ lattice cells to achieve different confinement ratios $2R/W$. The stream-flow (x) direction is resolved with $1024 \le L_{x} \le 1356$ lattice cells, depending on the droplet elongation properties, while the transverse-flow (y) direction is resolved with $128$ lattice cells. Periodic conditions are applied in the stream-flow and in the transverse-flow directions. The droplet is subjected to a linear shear flow $u_x=\dot{\gamma} z$, $u_y=u_z=0$, with the shear introduced with two opposite velocities in the stream-flow direction ($u_x(x,y,z=W)=-u_x(x,y,z=0)=U_w$) at the upper ($z=W$) and lower wall ($z=0$).\\
In Figs.~\ref{fig:1} and \ref{fig:2} we show the time history for droplets in post-critical conditions at changing confinement and viscoelasticity. Fig.~\ref{fig:1} refers to cases with lower confinement ratio and $\mbox{De}=2.0$ including matrix and droplet viscoelasticity. For each case we consider three representative snapshots showing (i) the initial droplet deformation (ii) the droplet deformation prior to break-up (iii) and the droplet in post break-up conditions. We use the droplet emulsion time $\tau_{em}$ \eqref{emulsiontime} as a unit of time. For the Newtonian case we find $\mbox{Ca}_{\mbox{\tiny{cr}}} = 0.34$, which is different from the usual unconfined result $\mbox{Ca}_{\mbox{\tiny{cr}}}=0.43$~\cite{Grace,Janssen10}. This can be attributed to the finite Reynolds number ($\mbox{Re} \approx 0.5$) of our simulations~\cite{RenardyCristini01}. This fact said, we observe that both droplet and matrix viscoelasticity do not have an important influence on the critical Capillary numbers for break-up. In the case of droplet viscoelasticity we find a small stabilization that increases the critical Capillary number by some percent; wheres matrix viscoelasticity is not producing any visible effect on the critical Capillary number. It must be emphasized that we tuned the polymeric viscosity in such a way to reproduce always a unitary viscosity ratio between the droplet phase and the matrix phase. Fig.~\ref{fig:2} is essentially the counterpart of Fig.~\ref{fig:1} for an increased confinement ratio. A series of hints are given by the visual inspection of the droplet shapes and the associated critical Capillary numbers. First, the confinement ratio is already large enough to stabilize long droplet shapes in the Newtonian case, thus triggering the emergence of {\it triple break-up}~\cite{Janssen10}. Such droplet shapes would be unstable in unconfined flows: confinement makes them stable and the droplet can sustain larger Capillary numbers before break-up. Break-up mechanism itself changes, as the droplet can reach a minimum length at which a Rayleigh-Plateau instability~\cite{Janssen10} develops at the interface and breaks the droplet in equally sized daughter droplets (Panel (c)).  This fact is known from the literature~\cite{Janssen10,Cardinaels11} and lends further support~\cite{SbragagliaGuptaScagliarini} to the validity of the numerical approach. Second, and more interestingly, the role of matrix and droplet viscoelasticity seems opposite. Droplet viscoelasticity reduces droplet elongation and higher Capillary numbers are needed to break the droplet. The break-up process still leads to the formation of multiple neckings but the degree of monodispersity of the resulting daughter droplets gets affected at the Deborah number studied~\cite{SbragagliaGuptaPRE}. On the other hand, we observe that matrix viscoelasticity completely suppresses the formation of multiple neckings and the break-up process looks much more similar to the unbounded case. We remark that the finite extensibility parameter has been kept fixed to $L^2=100$. In another study~\cite{SbragagliaGuptaPRE}, we investigated systematically the importance of the finite extensibility parameter for the case of droplet viscoelasticity. It has been found that increasing $L^2$ could lead to a situation where elongated droplet shapes cannot not be stable anymore due to the net increase of the polymer elongational viscosity, which actually increases at increasing $L^2$. In such a case, also with droplet viscoelasticity the critical Capillary number decreases with respect to the Newtonian case. 

\begin{figure}[t!]
\begin{center}
\includegraphics[scale=0.60]{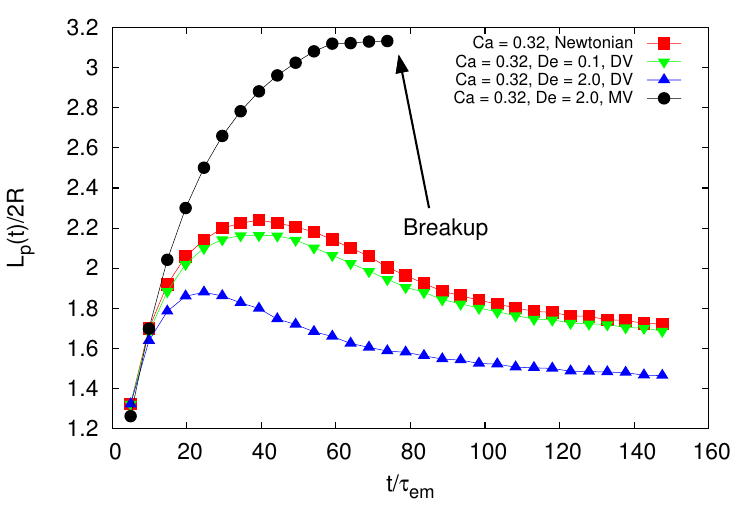}
\put(-180,128){ {\small { (a) } } }
\includegraphics[scale=0.60]{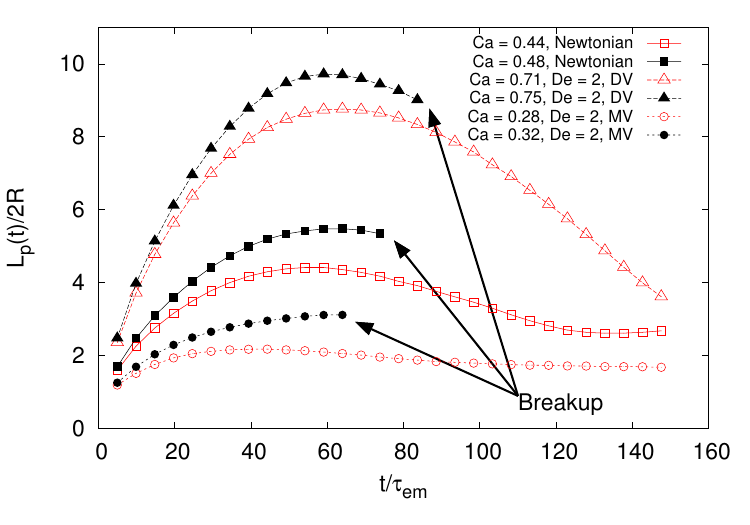}
\put(-180,128){ {\small { (b) } } }
\end{center}
\caption{Evolution of the dimensionless droplet length after the startup of a shear flow for various Capillary numbers and Deborah numbers for a fixed confinement ratio $2R/W = 0.78$ and finite extensibility parameter $L^2=10^2$. Since the shape of highly deformed droplets may deviate from an ellipsoid, we estimated the droplet elongation from the projection of the droplet length ($L_p$) in the velocity direction. The viscosity ratio between the droplet phase and the matrix phase is kept fixed to $\lambda=\eta_d/\eta_M=1$, independently of the degree of viscoelasticity. Similarly to Figs.~\ref{fig:1}-\ref{fig:2}, we use the droplet emulsion time $\tau_{\mbox{\tiny{em}}}$ (see Eq.~\eqref{emulsiontime}) as a unit of time.}
\label{fig:Lp}
\end{figure}

\begin{figure}[t!]
\begin{center}
\includegraphics[scale=0.60]{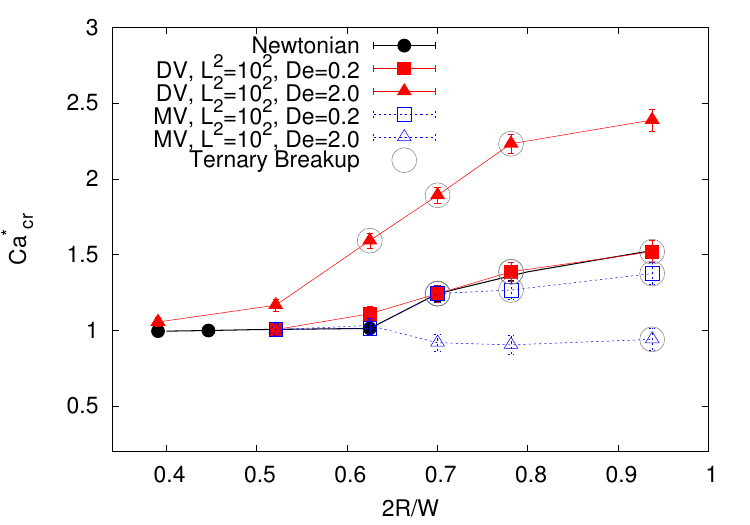}
\put(-30,130){ {\small { (a) } } }
\includegraphics[scale=0.60]{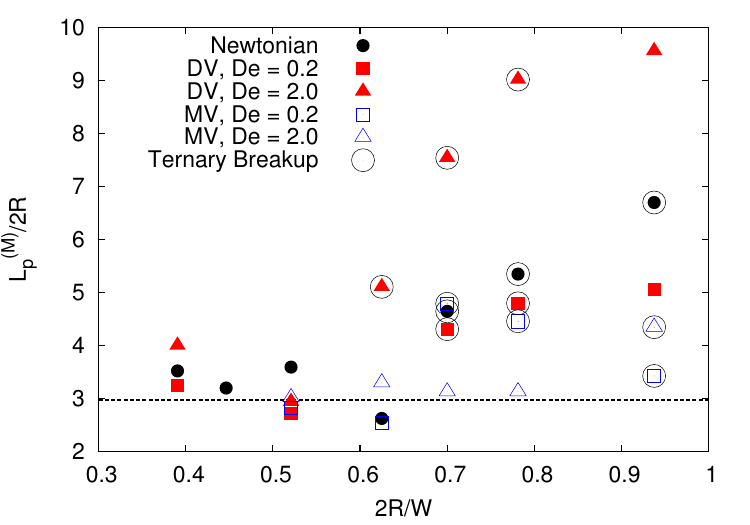}
\put(-188,130){ {\small { (b) } } }
\end{center}
\caption{Panel (a): Critical Capillary number for break-up as a function of confinement ratio for systems with finite extensibility parameter $L^2 = 10^2$. The critical Capillary number has been normalized to the value of the Capillary number in the unbounded case which is essentially the same for Newtonian and viscoelastic cases. The viscosity ratio between the droplet phase and the matrix phase is kept fixed to $\lambda=\eta_d/\eta_M=1$ in all cases. Different Deborah numbers are considered, by changing the polymer relaxation time in equations \eqref{NS}-\eqref{FENEb}. We consider both the cases of droplet viscoelasticity (DV) and matrix viscoelasticity (MV). Black open circles indicate situations where multiple necking occur. Panel (b): data analyzed in Panel (a) are reported in terms of the dimensionless maximum elongation of the droplet $L^{(M)}_p/2R$.}
\label{fig:Cacrandviscousratio}
\end{figure}

In Fig.~\ref{fig:Lp} we show the dimensionless droplet elongation $L_p/2R$ as a function of time for several values of $\mbox{De}$ and $\mbox{Ca}$, at fixed confinement ratio $2R/W=0.78$, comparing both matrix and droplet viscoelasticity with the Newtonian case. Since the shape of highly deformed and confined droplets deviates from an ellipsoid, we estimated the droplet elongation from the projection of the droplet length ($L_p$) in the velocity direction. In Panel (a) of Fig.~\ref{fig:Lp} we report results for a given Capillary number $\mbox{Ca}=0.32$. If compared with the Newtonian case, the maximum elongation of the droplet is suppressed in the case of droplet viscoelasticity, while is enhanced in the case of matrix viscoelasticity, which is a signature that the critical Capillary number for matrix viscoelasticity is smaller than the Newtonian one. This happens for Deborah number just above unity, whereas results with small Deborah numbers are actually very close to the Newtonian case. In Panel (b) we report the pre-critical and post-critical time history for both Newtonian and viscoelastic droplets. We notice that the maximum dimensionless elongation achieved before break-up, $L^{(M)}_p/2R$, is larger for the Newtonian case compared to the case of matrix viscoelasticity, which indeed breaks at smaller Capillary number.  On the other hand, the case with droplet viscoelasticity achieves a maximum elongation before break-up that is roughly doubled with respect to the Newtonian case. Also the critical Capillary number is roughly doubled (see Fig.~\ref{fig:Cacrandviscousratio}).

Panel (a) of Fig.~\ref{fig:Cacrandviscousratio} summarizes and extends the findings of Figs. \ref{fig:1}-\ref{fig:Lp} to other confinement ratios and degrees of viscoelasticity in both the matrix and droplet phases. We report data for weakly viscoelastic systems ($\mbox{De}=0.2$) and also data with Deborah number just above unity ($\mbox{De}=2.0$). As already noticed elsewhere~\cite{SbragagliaGuptaPRE}, for Newtonian droplets the role of confinement is almost insignificant up to $2R/W=0.625$, whereas for larger confinement ratios a monotonous increase of $\mbox{Ca}_{\mbox{\tiny{cr}}}$ is observed. The emergence of the up-turn in $\mbox{Ca}_{\mbox{\tiny{cr}}}$ is intimately connected to the change of the break-up mechanism.  For small Deborah numbers the curve $\mbox{Ca}_{\mbox{\tiny{cr}}}$ {\it vs.} $2R/W$ does not change much for droplet viscoelasticity, whereas some decrease in $\mbox{Ca}_{\mbox{\tiny{cr}}}$ can be readily seen for the case of matrix viscoelasticity. The black open circles indicate situations where ternary break-up is observed. Actually, the tendency of viscoelasticity to alter and change the stability properties of confined droplets is perceptibly more pronounced in the case of matrix viscoelasticity than droplet viscoelasticity. For Deborah number just above unity ($\mbox{De}=2.0$) both matrix and droplet viscoelasticity alter significantly the critical Capillary number at break-up and the changes are more pronounced and amplified at larger confinement ratios. In Panel (b) of Fig.~\ref{fig:Cacrandviscousratio}, we report the maximum dimensionless elongation of the droplet, $L^{(M)}_p/2R$, as a function of the confinement ratio. It is clear that the trends for $\mbox{Ca}_{\mbox{\tiny{cr}}}$ and $L^{(M)}_p/2R$ are quite similar.\\

\section{Conclusions}

The effects of viscoelasticity and geometrical confinement on droplet break-up in a shear flow have been studied. Our analysis strongly benefited of numerical simulations, where we could model immiscible fluids in the presence of viscoelastic behaviour in either the droplet or the matrix phase. Numerical simulations offer great flexibility and easy access to hydrodynamic quantities and droplet interface dynamics, thus revealing particularly suited for the study at hand. We have found that the effect of viscoelasticity is rather insignificant in unbounded cases, whereas it gets amplified with confinement. In particular, viscoelasticity reduces the capability of micro-confined shear flows to generate monodisperse emulsions. This is a feature that we found in a previous study on droplet viscoelasticity~\cite{SbragagliaGuptaPRE} and we confirm in presence of matrix viscoelasticity. At small Deborah number, the tendency of viscoelasticity to alter and change the stability properties of confined droplets is more pronounced in the case of matrix viscoelasticity, if compared to the case of droplet viscoelasticity.\\
For future research it would be extremely interesting to repeat some of the numerical simulations at changing the polymer concentration and/or at changing the geometry of the system (i.e. T-shaped channels, flow-focusing devices) to reveal other interesting features on the dynamics of viscoelastic fluids in complex geometries.

\section{Acknowledgments}

We are particularly grateful to F. Bonacccorso for technical support. We kindly acknowledge funding from the European Research Council under the European Community's Seventh Framework Programme (FP7/2007-2013) / ERC Grant Agreement  N. 279004. A. Gupta acknowledges R. Pandit and S. S. Ray for fruitful discussions during his visits to IISc Bangalore (February 2014, August 2014) and ICTS-TIFR Bangalore (October 2013).


\begin{thebibliography}{}

\bibitem{Christophher07} G.F. Christopher \& S.L. Anna {\it J Phys D Appl Phys} 2007 {\bf 40}:R319--R336

\bibitem{Larson} R.G. Larson, {\it The Structure and Rheology of Complex Fluids}, Oxford University Press, New York (1999)

\bibitem{Taylor34} G.I. Taylor, {\it Proc Roy Soc A} 1932; {\bf 138}:41--8.

\bibitem{Grace} H. P. Grace, {\it Chem. Eng. Commun.} 1982; {\bf 14}, 225 


\bibitem{Stone} H. A. Stone, {\it Annu. Rev. Fluid Mech.} 1994; {\bf 26}, 65 

\bibitem{Shapira} M. Shapira \& S. Haber, {\it Int J Multiph Flow} 1990;{\bf 16}:305--21

\bibitem{Sibillo06} V. Sibillo, G. Pasquariello, M. Simeone, V. Cristini \& S. Guido, {\it Phys Rev Lett} 2006; {\bf 97}, 054502

\bibitem{Vananroye07} A. Vananroye,  P. Van Puyvelde \& P. Moldenaers, {\it J. Rheol.} 2007; {\bf 51}, 139--153

\bibitem{Janssen10} P. J. A. Janssen, A. Vananroye, P. Van Puyvelde, P. Moldenaers \& P. D. Anderson, {\it J Rheol} 2010; {\bf 54}, 1047--1060 


\bibitem{Renardy07} Y. Renardy, {\it Rheol Acta} 2007; {\bf 46}, 521--529 

\bibitem{RenardyCristini01} Y. Renardy \& V. Cristini {\it Phys. Fluids} 2007; {\bf 13}, 7--13 

\bibitem{Cardinaels09}  R. Cardinaels, K. Verhulst \& P. Moldenaers, {\it J. Rheol} 2009; {\bf 53}, 1403--1424 

\bibitem{Minale10} M. Minale, S. Caserta \& S. Guido, {\it Langmuir} 2010; {\bf 26}, 126--132 

\bibitem{Cardinaels11} R. Cardinaels \& P. Moldenaers, {\it Microfluid Nanofluid} 2011; {\bf 10}, 1153--1163


\bibitem{bird} R. B. Bird, R. C Armstrong \& O. Hassager, {\it Dynamics of polymeric liquids}, J. Wiley \& sons (1987).

\bibitem{Onishi2} J. Onishi, Y. Chen, H. Ohashi, {\it Physica A} 2006; {\bf 362} 84--92

\bibitem{Xi99} H. Xi \& C. Duncan, {\it Phys. Rev. E.} 1999; {\bf 59}, 3022--3037

\bibitem{SC1} X. Shan \& H. Chen,  {\it Phys. Rev. E} 1993; {\bf 47}, 1815

\bibitem{CHEM09} R. Benzi, M. Sbragaglia, S. Succi, M. Bernaschi \& S. Chibbaro, {\it J. Chem. Phys.} 2009; {\bf 131}, 104903 

\bibitem{Komrakovaa13} A. E. Komrakovaa, O. Shardta, D. Eskinb \& J.J. Derksen , {\it International Journal of Multiphase Flow} 2013; {\bf 59}, 24--43

\bibitem{VanDerSman08} R.G.M. Van der Sman \& S. van der Graaf, {\it Comput. Phys. Commun.} 2008; {\bf 178}, 492--504

\bibitem{Liuetal12} H. Liu, A. J. Valocchi \& Q. Kang, {\it Phys. Rev. E} 2012; {\bf 85}, 046309

\bibitem{Malaspinas10} O. Malaspinas, N. Fietier \& M. Deville, {\it Jour. Non Newt. Fluid Mech} 2010; {\bf 165}, 1637--1653 

\bibitem{Herrchen} M. Herrchen \& H. C. Ottinger, {\it J. Non-Newtonian Fluid Mech.} 1997; {\bf 68}, 17--42 

\bibitem{Lindner03} A. Lindner, J. Vermant, D. Bonn, {\it Physica A} 2003; {\bf 319}, 125--133 


\bibitem{SbragagliaGuptaScagliarini} A. Gupta, M. Sbragaglia and A. Scagliarini, {\it arXiv/1406.2686} (2014)

\bibitem{SbragagliaGuptaPRE} A. Gupta and M. Sbragaglia, {\it Phys. Rev. E} {\bf 90}, 023305 (2014)


\end{thebibliography}
\end{document}